\documentclass[12pt]{article}
\usepackage[utf8]{inputenc}
\usepackage[english]{babel}
\usepackage{ragged2e}

\usepackage{scrextend}
\deffootnote[1em]{1em}{1em}{\textsuperscript{\thefootnotemark}\,}

\usepackage[margin=1in]{geometry} 
\usepackage{setspace}            
\usepackage{graphicx}
\usepackage{adjustbox}
\usepackage{caption}            
\usepackage{rotating}           

\usepackage{booktabs}  
\usepackage{threeparttable}    
\usepackage{dcolumn}    
    \newcolumntype{d}[1]{D{.}{.}{#1}}

\usepackage{natbib}
\usepackage[colorlinks=true, allcolors=blue, backref=page]{hyperref}

\usepackage{amssymb, amsfonts, amsmath}
\usepackage{bm}

\usepackage{longtable}

\usepackage{comment}

\let\estinput=\input 

\newcommand{\estwide}[3]{
		\vspace{.75ex}{
			\begin{tabular*}
			{\textwidth}{@{\hskip\tabcolsep\extracolsep\fill}l*{#2}{#3}}
			\toprule
			\estinput{#1}
			\\ \bottomrule          
			\addlinespace[.75ex]
			\end{tabular*}
			}
		}	

\newcommand{\estauto}[3]{
		\vspace{.75ex}{
			\begin{tabular}{l*{#2}{#3}}
			\toprule
			\estinput{#1}
			\\ \bottomrule          
			\addlinespace[.75ex]
			\end{tabular}
			}
		}

\title{\Large The Effect of External Debt on Greenhouse Gas Emissions\thanks{The opinions expressed herein are solely those of the authors and do not necessarily represent the view(s) of affiliated institutions.}
\vspace{-1.2em}}

\author{
    Jorge Carrera\thanks{National Scientific and Technical Research Council (CONICET), and University of La Plata, Argentina}
    \and
    Pablo de la Vega\thanks{Instituto de Investigaciones Económicas, Facultad de Ciencias Económicas, Universidad Nacional de La Plata, Argentina. Corresponding author: \href{mailto:delavegapc@gmail.com}{delavegapc@gmail.com}}
}

\date{\large  Last updated: August 5, 2023}

\date{\today} 

\begin{document}

\maketitle

\begin{abstract}
\noindent 
We estimate the causal effect of external debt on greenhouse gas emissions in a panel of 78 emerging market and developing economies over the 1990-2015 period. Unlike previous literature, we use external instruments to address the potential endogeneity in the relationship between external debt and greenhouse gas emissions. Specifically, we use international liquidity shocks as instrumental variables for external debt. We find that dealing with the potential endogeneity problem brings about a positive and statistically significant effect of external debt on greenhouse gas emissions: a 1 percentage point (pp.) rise in external debt as a percentage of GDP causes, on average, a 0.5\% increase in greenhouse gas emissions. One possible mechanism of action could be that, as external debt increases, governments are less able to enforce environmental regulations because their main priority is to increase the tax base to pay increasing debt services or because they are captured by the private sector who owns that debt and prevented from tightening such regulations.

\vspace{-1em}
\end{abstract}

\footnotesize{\textbf{Keywords}: external debt, GHG emissions, instrumental variables, panel data. \vspace{-1em}}

\footnotesize{\textbf{JEL}: C23, C26, F64, H63.}

\newpage




\newpage
\section{Introduction}

Human-induced climate change due to environmental degradation is probably the most serious problem that humankind is facing. The concentration of greenhouse gases (GHG)—particularly, carbon dioxide (CO$_{2}$)—in the atmosphere has risen at a tremendous pace in recent decades \citep{IPCC2021}. Because of the threat they pose to climate stability, biodiversity, and economic development, there has been a great deal of academic and political interest in the determinants of GHG emissions. This knowledge is crucial to design policies towards sustainable development.

Although the literature on GHG emissions determinants is wide and long dating, the study of the relationship between external debt and GHG emissions has recently attracted scholarly attention. These works are aimed at determining the relationship between external debt and GHG emissions in countries such as China, India, and other emerging market and developing economies (EMDEs). This question is crucial since, over the last decades, there has been a widespread build-up of debt in almost every country in the world, and the public sector has led this process \citep{antoniades2018, yared2019, morenobadia2020, kose2020}. Moreover, the nexus between external debt and GHG emissions is of particular importance in EMDEs. These countries are still catching up in terms of economic development and are therefore expected to continue to grow for several decades. Debt issuance is key to catalyze such process. 

Our main research question is what is the effect of external debt on GHG emissions? We only found a few papers that address this relationship, most of which deal with a single country \citep{katircioglu2018, bese2021a, bese2021b, bese2022, Bachegour2023}. As regards our work, it is more aligned with \cite{akam2021}, who use a panel of thirty-three heavily indebted poor countries. However, we expand the scope of analysis by using a wide panel of seventy-eight EMDEs. In addition, unlike previous literature, we use external instruments to deal with potential endogeneities in the relationship between external debt and GHG emissions. In particular, we exploit the exposure to global push factors of international monetary liquidity \citep{reinhart2009, forbes2012, rey2015} as an exogenous variation in external debt.

When taking into account the potential endogeneity problem, we find a positive and statistically significant effect of external debt on GHG emissions. A 1 percentage point (pp.) rise in external debt causes, on average, a 0.5\% increase in GHG emissions. 
In exploring possible a mechanism of action, we find that external debt is negatively related to an indicator of policies associated with environmental sustainability. This may suggest that when external debt increases, governments are less able to enforce environmental regulations because their main priority is to increase the tax base to pay increasing debt services or because they are captured by the private sector and prevented from tightening such regulations, and therefore could explain the positive associacion between external debt and environmental degradation.

Considering the key role that external debt should play in financing the green transition, our results suggest a paradox, as higher external debt has so far been contributing to environmental damage. If EMDEs are expected to fulfill their global and individual environmental commitments, the effect of external debt on GHG emissions should be reversed.

The remaining part of the paper is as follows. In Section \ref{sec:litreview}, we provide a literature review. In Section \ref{sec:strategy}, we present the empirical strategy, whereas in Sections \ref{sec:baseline} we report the results. Finally, conclusion is reached in Section \ref{sec:conclusion}.




\section{Literature Review} \label{sec:litreview}

The empirical study of GHG emissions determinants has been traditionally framed into the Environmental Kuznets Curve hypothesis since the early nineties. Briefly, this hypothesis states that environmental degradation is an increasing function of economic development up to a critical income level when higher economic incomes are negatively associated with environmental damage following an inverted U-shape pattern. Extensive literature reviews can be found in \cite{shahbaz2019}, \cite{sanchez2019}, \cite{jiang2019}, \cite{pincheira2021}.   

Notwithstanding, the empirical assessment of the relationship between external debt and GHG emissions is highly new and scarce. These recent papers were aimed at estimate the relationship between external debt and GHG emissions GHG emissions in countries such as China, India, and other EMDEs. By using a vector error correction model, \cite{katircioglu2018} do not find any evidence for the effect of external debt on CO$_{2}$ emissions in Turkey during the 1960-2013 period. In the case of China, \cite{bese2021a} use ARDL and nonlinear ARDL models and find a significant positive effect of external debt on CO$_{2}$ emissions for the 1978-2014 period. By using the same methodologies, \cite{bese2021b} analyze the effect of external debt on different types of emissions in India between 1971 and 2012. They find a positive effect of external debt on emissions of CO$_{2}$, methane, and those from gaseous and solid fuel consumption. Similarly, \cite{bese2022} find an inverted U-shaped relationship between external debt and CO$_{2}$ emissions in Turkey over the period 1970-2016. The authors hypothesize that this is because infrastructure investments may have contributed to reducing environmental pollution. Moreover, \cite{Bachegour2023} estimate an ARDL model over the period 1984-2018 and find that external debt reduces carbon dioxide emissions in Morocco. In a panel of 33 heavily indebted poor countries (HIPC) from 1990 to 2015, \cite{akam2021} find a positive, though not significant, association between external debt and CO$_{2}$ emissions. 

There is a main plausible channel through which external debt could affect GHG emissions. Economic growth driven by external debt, e.g., due to investments \citep{shabbir2013, zaman2014, keshmeer2021}\footnote{It is worth bearing in mind that the nexus between debt and growth has been extensively studied, but even today no definitive conclusion on the direction of causality between the two variables has been made \citep{modigliani1961, bernheim1987, lyoha1999, aschauer2000, reinhart2010, carrera2021, law2021}}, could induce an increase in energy consumption and thus environmental pollution.\footnote{\cite{katircioglu2018} and \cite{akam2021} discuss other plausible channels such as external debt financing energy imports in energy deficit economies. However, from a money fungibility perspective, this may be hard to assess and we consider the economic growth channel to include this case well.} This is especially true if investments are directed to GHG-intensive sectors of activity, such as real estate, manufacturing and construction, or if the additional financing generates a higher level of economic activity and consumption of energy supplied by emission-intensive energy sources, such as fossil fuels. However, environmental damage due to higher economic growth and energy consumption financed with external debt could be reduced if that financing is associated with the use of cleaner energies \citep{shafiei2014, hanif2018, nguyen2019, kahia2019, khattak2020}.\footnote{This is in fact related to the increasing, although yet low, relevance of green bonds \citep{ehlers2017, flammer2020, tolliver2020}}



\section{Empirical Strategy} \label{sec:strategy}

In order to analyze the relationship between external debt and GHG emissions, we estimate the following equation:
\begin{equation}
    CO_{2_{i,{t}}} = \alpha_{i,{t}} + \beta ED_{i,{t}} + \theta_0 Y_{i,{t}} + \theta_1 Y_{i,{t}}^2 + \delta X_{i,{t}} + \gamma_t + \mu_i + \epsilon_{i,{t}}  \label{eq:eq1}
\end{equation}
where $CO_{2_{i,{t}}}$ is the log of per capita carbon dioxide emissions (thousand metric tons); $ED_{i{t}}$ is the ratio of external debt over GDP; $Y_{i,{t}}$ is the log of the GDP per capita, which is included in a linear and quadratic way to test for the existence of the Environmental Kuznets Curve. We use CO$_{2}$ emissions because they account for the largest share of global GHG emissions since 1970 \citep{IPCC2021, olivier2020}. $X_{i{t}}$ is a vector of control variables that is common place in the Environmental Kuznets Curve literature including as follows:

\begin{itemize}
    \item CO$_{2_{i,{t-1}}}$ to control for the inertia in the GHG emissions. It is expected to be positive, suggesting rigidities in the energy consumption matrix composition and thus a path-dependence trajectory \citep{zilio2014}.
    \item The productive structure proxied by the share of agricultural and industry value-added in GDP. One of the traditional explanations of the Environmental Kuznets Curve lies in the fact that the transition from an agricultural to an industrial economy would lead to higher environmental degradation since the latter is relatively more pollutant than the former. In other words, GHG emissions would start to significantly increase as an economy enter the industrialization phase. Previous findings show ambiguous results for the share of agricultural value-added in GDP \citep{liu2017, gokmenoglu2019, aziz2020, ridzuan2020}. In addition, other authors suggest that a higher industry share could be negatively related to emissions due to the development and absorption of efficient and environmentally friendly technologies \citep{dogan2020}.
     \item A human capital index as a proxy for the environmental quality demand. A society with a high human capital level is likely to be more aware of the negative consequences of environmental degradation and, thus, would demand environmental quality in the form of regulations, carbon taxes, environmentally friendly technologies, etc. \citep{pata2021}. \cite{poudel2009} find that the illiteracy rate, defined as the percentage of the population aged 15 years who are not able to read, is negatively correlated with carbon emissions. \cite{sarkodie2020} find that increasing human capital positively affects environmental degradation in China, whereas \cite{pata2021} find an opposite result. \cite{hassan2019} find insignificant effects of human capital on the ecological footprint in Pakistan.
     \item The share of renewable energy consumption in total energy consumption and the share of fossil fuel energy consumption in total energy consumption to test for energy consumption structure. GHG is expected to decrease, with the former and increase with the latter \citep{allard2018, sanchez2019, dogan2020}. The omitted share of energy is nuclear, which may be considerable only for some countries. We will check how its inclusion affects the results in Section \ref{sec:robustness}     
     \item The energy use (kg of oil equivalent per capita), which is expected to be positively related to GHG emissions \citep{nasir2011, almulali2013, sirag2018, hoangphong2019}. 
     \item The forest area (\% of land area) as a proxy for the availability of the natural resources given that forests are an important part of the biocapacity, the offsetting force of the environmental damage. \cite{hassan2019} find that biocapacity does not affect the ecological footprint in Pakistan. \cite{poudel2009} neither find a significant effect in 15 Latin American countries.
     \item The Domestic credit to the private sector (\% of GDP) as a proxy of financial development. It is expected to be positively related to GHG emissions because of the availability to finance projects and enterprises which, in turn, increase energy consumption \citep{javid2016, hoangphong2019, shahbaz2020}.
     \item Financial and Trade Globalization Indexes. The expected sign is ambiguous for both variables as predicted by theory \citep{grossman1991}. According to \cite{shahbaz2013, shahbaz2015, shahbaz2017a}, globalization could increase pollution if trade and capital flows induce an economic expansion \citep{dinda2008, sirag2018, hoangphong2019} and, especially, if investments are directed to emission-intensive production. However, if globalization is associated with the diffusion of efficient and environmentally friendly technologies and institutions, it could negatively affect GHG emissions \citep{runge1994, wheeler2000, jayadevappa2000, liddle2001, cole2006}. Furthermore, the ``Pollution Haven'' Hypothesis suggests a process of relocation of polluting industries from developed economies with tight environmental regulation to countries beginning the industrialization phase and few environmental regulations \citep{copeland2004}. In consequence, trade would be positively associated with GHG emissions in low and middle-income countries but negatively related to high-income countries \citep{zilio2014, allard2018, sanchez2019}. \cite{managi2004} finds that trade liberalization is positively associated with carbon emissions in a panel of sixty-three developed and developing countries over the 1960-1999 period. Notwithstanding, previous literature also finds non-significant results for the trade openness variable \citep{lee2009, you2015}. Recent works deal with other measures of globalization. \cite{shahbaz2013} and \cite{shahbaz2015} find that the KOF globalization index is negatively correlated with CO$_{2}$ emissions in Turkey and India, respectively. Furthermore, \cite{lee2014} find that the KOF index negatively affects carbon emissions for a panel of 255 countries from 1980 to 2011. However, \cite{shahbaz2017a} and \cite{shahbaz2017b} find a positive relationship for Japan and twenty-five developed economies, respectively.
\end{itemize}

Finally, $\gamma_t$ and $\mu_i$ are time and country fixed effects, respectively. The dataset comprises annual data for 78 EMDEs from 1990 to 2015. The list of countries included, definition and source of data, and some descriptive statistics of both the dependent variables and the independent variables are shown in Appendix \ref{sec:appendix}.

The coefficient of interest is $\beta$. Although the reverse causality is unlikely\footnote{Notwithstanding, having climate change-related policies in place is increasingly becoming a prerequisite for governments and firms to receive funding \citep{wire2021, rankin2021}. Therefore, it is likely that reverse causality may be also an issue in the future.}, the relationship between external debt and GHG emissions could be driven by a third factor, such as country-specific shocks or fundamentals, which rise the possibility of endogeneity problems. To address this potential issue, we instrument $ED_{i{t}}$ with a proxy for the country i’s exposure to international financial liquidity shocks defined as follows:
\begin{equation}
    z_{i,{t}}  = i^*_t \overline{KOPEN}_{i}  \label{eq:eq2}
\end{equation}
\begin{equation}
    \overline{KOPEN}_{i} = \frac{1}{t} \sum_t KOPEN_{i,t} = \frac{1}{t} \sum_t \sum_j \omega_{i,j} KOPEN_{j,i,t}  \label{eq:eq3}
\end{equation}
where $i^*$ is the Federal Funds Effective Rate as a proxy of international monetary liquidity, and $\overline{KOPEN}_{i}$ is the time-average of the weighted average Chinn-Ito de-jure financial openness index \citep{chinnito2006} of $i’s$ neighbors countries using geographical distance as weights ($\omega_{i,j}$).\footnote{In order to increase sample size, for each country $i$, we use all $j$ countries with information on financial openness and calculate weights as the inverse of the simple distance between most populated cities, which is constant, obtained from CEPII GeoDist Database \citep{mayer2011}} In this way, we exploit the regional exposure to global push factors \citep{reinhart2009, forbes2012, rey2015, aizenman2016, miranda2020}, independent of country-specific shocks and fundamentals, as an exogenous variation in external debt.\footnote{This instrument is similar to the one used in \cite{habib2017} for estimating the causal impact of real exchange rate on economic growth. However, they use global capital flows as a proxy of international liquidity and the $i$ country’s financial openness instead of the time-averaged financial openness of the $i’s$ neighbors countries.} Financial openness measures may be correlated with other government policies—such as those oriented to industrial production, energy use structure, and environmental protection, among others—that also affect GHG emissions. For this reason, we use a weighted average financial globalization index of country $i’s$ neighboring countries to proxy for country $i’s$ exposure to international financial liquidity. Moreover, we time-average $KOPEN_{i,{t}}^j$ so the only time-series variation in the instrument is from $i^*$, which is arguably exogenous. A caveat should be lodged on financial flows other than external debt, such as FDI, since they may induce economic growth or finance investment that, in turn, affect GHG emissions. This would represent a violation of the exclusion restriction. In consequence, we also control for GDP growth and net capital inflows.  

The relevance of the instruments will be assessed through the test for under-identification and the Anderson-Rubin Wald test for weak identification. The first is an LM test to determine whether the excluded are correlated with the endogenous regressors. The null hypothesis is the absence of correlation. The second tests the statistical significance of the coefficient ($\beta$) associated with the endogenous variable in the presence of potentially weak instruments. The null hypothesis is $\beta$=0 and a confidence interval $L$\% is the set of all $\beta_0$ values such that the null hypothesis cannot be rejected at the (100-$L$)\% significance level \citep{stock2005, finlay2013}. In addition, there is the usual "rule of tumb" that the first-stage F-statistic should be larger than 10. In the special case of single endogenous regressor, the Kleibergen-Paap robust rk Wald F statistic is equivalent to the F-test from first-stage regression, while the Kleibergen-Paap robust rk LM statistic for identification is equivalent to the LM test of joint significance of excluded instruments in first-stage regression and to the LM test of an excluded instrument for redundancy.

It is worth noting that the instrumental variable method only identifies a local average treatment effect, i.e., an effect just for those affected by the instrument \citep{imbens1994}. Thus, in our setting, we estimate the effect of external debt on GHG emission for those affected by international monetary liquidity shocks. 



\section{Results} \label{sec:baseline}

The results of estimating equation \ref{eq:eq1} are shown in Table \ref{tab:table_lnco2_emi_tot}. When using the fixed effects (FE) estimator, we find a not significant effect of external debt on GHG emissions. However, when we take into account the potential endogeneity problem, we find a positive and statistically significant effect of external debt on GHG emissions (column 2). In particular, a rise of 1 pp. in external debt causes, on average, a 0.5\% increase GHG emissions, other variables remain constant. These results suggest the presence of omitted variable bias in the FE estimates, i.e., there is at least one omitted factor that is negatively correlated with GHG emissions but positively with external debt (or vice versa).\footnote{Note that the FE estimates are the same if we include GDP growth and net capital inflows as in the IV estimates. These results are available upon request.} Indeed, the endogeneity issue may be behind previous findings of the non-significant effect of external debt on GHG emissions \citep{akam2021}. 

First stage results are presented in the third column of Table \ref{tab:table_lnco2_emi_tot}. As expected, we find that a negative shock on international liquidity is negatively associated with the ratio of external debt over GDP. In other words, countries borrow less facing an increase in the global cost of liquidity either because it is more difficult for them to roll-over existing debt or because it is more expensive to issue new debt. At the bottom of the table, we report statistical tests to evaluate the relevance of the instruments. In particular, we show the first stage F-statistic, the test for under-identification, and the Anderson-Rubin Wald test for weak identification. We reject the null of under-identification and find that the coefficient associated with the external debt is statistically significant in the presence of weak identification. The F-statistics is well above 10.

The control variables mostly exhibit the expected signs. The autoregressive parameter shows a high persistence. We find positive and negative coefficients for the linear and quadratic term of the GDP per capita, which is consistent with the Environmental Kuznets Curve hypothesis. The agriculture share in the value added is negatively correlated with GHG emissions, whereas the industry share is positively and significantly related with GHG emissions. We do not find any significant effect of the human capital index. The share of renewable energy consumption is not significantly associated with GHG emissions, whereas fossil fuel energy consumption and the energy use are positively associated with GHG emissions. The forest area is not significantly related to emissions. The credit to private sector, trade and financial globalization are negatively related with emissions. Finally, in relation to the additional controls included in the IV estimates, net capital inflows and GDP growth are positively associated with emissions.

\clearpage
\begin{table}[!htbp]
 \caption{Baseline results}\label{tab:table_lnco2_emi_tot}
    \centering
    \footnotesize
    \singlespacing
    \scalebox{0.85} {   
	  \setlength{\tabcolsep}{10pt} 
	  \estauto{table_lnco2_emi_tot.tex}{3}{c}
	  }
  \justify
 \scriptsize
 Notes: The table reports FE and IV estimates with robust standard errors (in parenthesis), including time and country fixed-effects. In columns 1 and 2, the dependent variable is the log of CO$_{2}$ emissions (thousand metric tons of CO2), whereas in column 3 is external debt. The external debt is instrumented by the exposure of country i’s to international liquidity shocks (see equation 2). *** p\textless0.01, ** p\textless0.05, * p\textless0.1. The sample period is 1991-2015. F first stage is the F statistic for the first stage of the instrumental variables estimates. The Kleibergen-Paap rk LM statistic test for the relevance of the instruments. The Anderson-Rubin Wald Test statistic test for the statistical significance of the main (beta) coefficient associated with the external debt in the presence of potentially weak instruments.
 \end{table}

\clearpage

\clearpage

\subsection{Additional control variables} \label{sec:robustness}

As a robustness tests, we include additional control variables that can affect GHG emissiones:
\begin{itemize}
    \item The share of urban population (\% of the total population). The findings of previous literature are mixed \citep{dogan2020}. According to theory, it is expected that urbanization positively affects emissions at the initial stages of development when people move from rural to urban areas and energy consumption increases. Indeed, some works find a positive effect on emissions \citep{hoangphong2019}. However, other papers find the opposite result \citep{poumanyvong2010, pachauri2008}.
    \item 	The share of alternative and nuclear energy (\% of total energy use). As it is cleaner in comparison with fuel fossil sources a negative relationship is expected \citep{toth2006, menyah2010}. This indicator complements the share of renewables used in the baseline estimates, which does not include nuclear energy.
\end{itemize}
The results are shown in Table \ref{tab:table_lnco2_emi_tot_morecontrols} and are almost the same as those in Table \ref{tab:table_lnco2_emi_tot}. Statistical tests support the relevance of the instrument used. The control variables also exhibit the same signs as those in Table \ref{tab:table_lnco2_emi_tot}. Both additional controls variables are not significantly correlated with GHG emissions.

\clearpage
\begin{table}[!htbp]
 \caption{Robustness tests: Additional controls} \label{tab:table_lnco2_emi_tot_morecontrols}
    \centering
    \footnotesize
    \singlespacing
    \scalebox{0.86} {   
	  \setlength{\tabcolsep}{10pt} 
	  \estauto{table_lnco2_emi_tot_morecontrols.tex}{3}{c} 
	  }
 \justify
 \scriptsize
 Notes: The table reports FE and IV estimates with robust standard errors (in parenthesis), including time and country fixed-effects. In columns 1 and 2, the dependent variable is the log of CO2 emissions (thousand metric tons of CO2), whereas in column 3 is external debt. The external debt is instrumented by the exposure of country i’s to international liquidity shocks (see equation 2). *** p\textless0.01, ** p\textless0.05, * p\textless0.1. The sample period is 1991-2015. F first stage is the F statistic for the first stage of the instrumental variables estimates. The Kleibergen-Paap rk LM statistic test for the relevance of the instruments. The Anderson-Rubin Wald Test statistic test for the statistical significance of the main (beta) coefficient associated with the external debt in the presence of potentially weak instruments. 
\end{table}


\clearpage

\clearpage
\subsection{Potential Mechanism}\label{sec:mechanism}

As discussed above, there is a plausible main channel through which external debt could affect GHG emissions. External debt-driven economic growth, e.g., due to investment, could increase energy consumption and, thus, environmental pollution. However, in our baseline estimates we find a positive effect of external debt on GHG emissions even controlling for GDP growth. Therefore, external debt must affect emissions through another mechanism. One possible mechanism could be that, when external debt increases, governments are less able to enforce environmental regulations because their main priority is to increase the tax base to pay increasing debt services or because they are captured by the private sector and are prevented from tightening such regulations. Similarly, \cite{Woolfenden2023} argues that debt is preventing the phase-out of fossil fuels in Global South countries. The pressure to repay debt forces them to continue to invest in fossil fuel projects to repay loans from richer countries and financial institutions.

To test our hypothesis, we estimate the relationship between external debt and the index of policies and institutions for environmental sustainability developed by the World Bank. This index measures the extent to which environmental policies promote the protection and sustainable use of natural resources and pollution management (1=low to 6=high). Descriptive statistics can be found in the Appendix \ref{sec:appendix}. This new dataset starts in 2005 and has a lower country coverage (52 countries).

As can be seen in Table \ref{tab:table_policy_enviro}, we find a negative relationship, which could serve as evidence of this hyphotesis, and therefore could explain the positive associacion between external debt and environmental degradation.
A 1 percentage point increase in the external debt to GDP ratio leads to 0.01 standard deviations reduction in the index of policies and institutions for environmental sustainability.

Since the sample size is reduced when using the World Bank index of policies and institutions for environmental sustainability, we re-estimate our baseline results (Table \ref{tab:table_lnco2_emi_tot}) over the overlapping sample in order to provide evidence regarding the generalizability of our findings. The results presented in Table \ref{tab:table_lnco2_emi_tot_overlapping_subsample} again suggest that our findings are robust, despite the reduced sample size. Moreover, the potential mechanism discussed above is a relevant explanation for the positive association between external debt and environmental degradation that we have found.

\clearpage
\begin{table}[!htbp]
 \caption{Potential Mechanism}\label{tab:table_policy_enviro}
    \centering
    \footnotesize
    \singlespacing
    \scalebox{0.9} {   
	  \setlength{\tabcolsep}{10pt} 
	  \estauto{table_policy_enviro.tex}{3}{c}
	  }
  \justify
 \scriptsize
 Notes: The table reports FE and IV estimates with robust standard errors (in parenthesis), including time and country fixed-effects. In columns 1 and 2, the dependent variable is an Index of policy and institutions for environmental sustainability (1=low to 6=high), whereas in column 3 is external debt. The external debt is instrumented by the exposure of country i’s to international liquidity shocks (see equation 2). *** p\textless0.01, ** p\textless0.05, * p\textless0.1. The sample period is 1991-2015. F first stage is the F statistic for the first stage of the instrumental variables estimates. The Kleibergen-Paap rk LM statistic test for the relevance of the instruments. The Anderson-Rubin Wald Test statistic test for the statistical significance of the main (beta) coefficient associated with the external debt in the presence of potentially weak instruments.  
 \end{table}

\clearpage
\begin{table}[!htbp]
 \caption{Baseline results in the overlapping subsample}\label{tab:table_lnco2_emi_tot_overlapping_subsample}
    \centering
    \footnotesize
    \singlespacing
    \scalebox{0.85} {   
	  \setlength{\tabcolsep}{10pt} 
	  \estauto{table_lnco2_emi_tot_overlapping_subsample.tex}{3}{c}
	  }
  \justify
 \scriptsize
 Notes: The table reports FE and IV estimates with robust standard errors (in parenthesis), including time and country fixed-effects. In columns 1 and 2, the dependent variable is the log of CO$_{2}$ emissions (thousand metric tons of CO$_{2}$), whereas in column 3 is external debt. The external debt is instrumented by the exposure of country i’s to international liquidity shocks (see equation 2). *** p\textless0.01, ** p\textless0.05, * p\textless0.1. The sample period is 1991-2015. F first stage is the F statistic for the first stage of the instrumental variables estimates. The Kleibergen-Paap rk LM statistic test for the relevance of the instruments. The Anderson-Rubin Wald Test statistic test for the statistical significance of the main (beta) coefficient associated with the external debt in the presence of potentially weak instruments.
 \end{table}


\clearpage
\section{Conclusion} \label{sec:conclusion}

We contribute to the recent study of the relationship between external debt on GHG emissions with causal evidence in a wide panel of countries. We estimate the impact of external debt on GHG emissions in a panel of 78 EMDEs from 1990 to 2015 and, unlike previous literature, we use external instruments to address potential endogeneity problems. Specifically, we use international liquidity shocks as instrumental variables for external debt. 

We find a positive and statistically significant effect of external debt on GHG emissions when we take into account the potential endogeneity problems. A 1 pp. rise in external debt causes, on average, a 0.5\% increase in GHG emissions. 
In exploring a possible mechanism of action, we find that external debt is negatively related to an indicator of policies associated with environmental sustainability. This may suggest that when external debt increases, governments are less able to enforce environmental regulations because their main priority is to increase the tax base or because they are captured by the private sector and prevented from tightening such regulations, and therefore could explain the positive associacion between external debt and environmental degradation. 

Our results point to significant negative environmental effects of external financing in EMDEs with several implications. First, this suggests that the endogeneity issue may be behind previous findings of the non-significant effect of external debt on GHG emissions \citep{katircioglu2018, akam2021}. On the contrary, our results are aligned with \cite{bese2021a} and \cite{bese2021b} who find a significant positive effect of external debt on CO$_{2}$ emissions in China and India, respectively.

Given the strong political intention to promote sustainable finance - as stated in different global policy forums such as the G20, FSB, IEA and COP26 - our results could serve as a reference to analyze how the relationship between external debt and greener finance should evolve in the future. According to the \cite{WEF2021}, massive financing from advanced economies, multilateral institutions and private markets to developing economies is needed to finance the transition to greener economies. However, if external debt plays the paradoxical role that we have found in our results, the outcome could be the opposite of that expected. If developing countries are expected to fulfill their global and individual environmental commitments, the effect of external debt on GHG emissions should be reversed.

In terms of limitations of our study, as stated before, the instrumental variable method only identifies a local average treatment effect, i.e., an effect just for those affected by the instrument \citep{imbens1994}. Thus, in our setting, we estimate the effect of external debt on GHG emission for those affected by international monetary liquidity shocks. 

\clearpage
\textbf{Declarations}

Funding: This research received no specific grant from any funding agency in the public, commercial, or not-for-profit sectors.

Conflict of Interest: We have no conflicts of interest to disclose.

Ethical approval: Not applicable.

Availability of data and materials: The datasets used and/or analysed during the current study are available from the corresponding author on reasonable request.

Authors' contributions:
\begin{itemize}
\item{JC: Conceptualization, Writing – original draft, Writing – review \& editing, Supervision, Project administration.} 

\item{PV: Methodology, Software, Formal analysis, Investigation, Writing – original draft, Writing – review \& editing, Visualization.}

\item{All authors read and approved the final manuscript.}
\end{itemize}


\clearpage

\bibliography{bibliography}  


\clearpage

\appendix
\section{Data description} \label{sec:appendix}

\setcounter{table}{0}
\renewcommand{\thetable}{A\arabic{table}}

The seventy-eight countries included in the dataset are listed below:
\begin{itemize}
    \item Africa (29): Algeria, Benin, Botswana, Cameroon, Congo, Congo, Dem. Rep., Cote d'Ivoire, Egypt, Ethiopia, Gabon, Gambia, Ghana, Kenya, Lesotho, Mauritius, Morocco, Mozambique, Niger, Nigeria, Senegal, South Africa, Sudan, Swaziland, Tanzania, Togo, Tunisia, Yemen, Zambia, Zimbabwe.
    \item Asia (22): Armenia, Bangladesh, Cambodia, China, Fiji, India, Indonesia, Iran, Jordan, Kyrgyzstan, Maldives, Mongolia, Myanmar, Nepal, Pakistan, Philippines, Russia, Sri Lanka, Syria, Tajikistan, Thailand, Vietnam.
    \item Europe (7): Albania, Bulgaria, Moldova, Romania, Serbia, Turkey, Ukraine.
    \item Latin America and the Caribbean (20): Argentina, Belize, Bolivia, Brazil, Colombia, Costa Rica, Dominican Republic, Ecuador, El Salvador, Guatemala, Guyana, Haiti, Honduras, Jamaica, Mexico, Nicaragua, Panama, Paraguay, Peru, Venezuela.
\end{itemize}

The classification based on GHG emissions level is as follows:
\begin{itemize}
    \item Low (24): Benin, Cameroon, Congo, Dem. Rep., Cote d'Ivoire, Ethiopia, Gambia, Ghana, Kenya, Mozambique, Niger, Nigeria, Senegal, Sudan, Tanzania, Togo, Zambia, Bangladesh, Cambodia, Myanmar, Nepal, Pakistan, Sri Lanka, Tajikistan, Haiti.
    \item Medium (26): Congo, Egypt, Eswatini, Lesotho, Morocco, Yemen, Zimbabwe, Armenia, Fiji, India, Indonesia, Kyrgyzstan, Philippines, Vietnam, Albania, Belize, Bolivia, Brazil, Colombia, Costa Rica, El Salvador, Guatemala, Honduras, Nicaragua, Paraguay, Peru.
    \item High (28): Algeria, Botswana, Gabon, Mauritius, South Africa, Tunisia, China, Iran, Jordan, Maldives, Mongolia, Russia, Syria, Thailand, Bulgaria, Moldova, Romania, Serbia, Turkey, Ukraine, Argentina, Dominican Republic, Ecuador, Guyana, Jamaica, Mexico, Panama, Venezuela.
\end{itemize}    


The definition and source of the variables used are indicated below:

\begin{itemize}
    \item CO$_{2}$ emissions (thousand metric tons of CO$_{2}$): Carbon dioxide emissions are those stemming from the burning of fossil fuels and the manufacture of cement. They include carbon dioxide produced during consumption of solid, liquid, and gas fuels and gas flaring. Source: World Bank WDI.
    \item External Debt (\% GDP): Total external debt is debt owed to nonresidents repayable in currency, goods, or services. Total external debt is the sum of public, publicly guaranteed, and private nonguaranteed long-term debt, use of IMF credit, and short-term debt. Short-term debt includes all debt having an original maturity of one year or less and interest in arrears on long-term debt. Source: International Debt Statistics.
    \item Index of policy and institutions for environmental sustainability: Measures the extent to which environmental policies promote the protection and sustainable use of natural resources and pollution management (1=low to 6=high). Source: World Bank's CPIA.    
    \item log of GDP per capita:	GDP per capita based on purchasing power parity (PPP) (constant 2017 international \$). Source: World Bank WDI.
    \item Agricultural, value added (\% of GDP):	Agriculture corresponds to ISIC divisions 1-5 and includes forestry, hunting, and fishing, as well as cultivation of crops and livestock production. Source: World Bank WDI.
    \item Industry, value added (\% of GDP):	It comprises value added in mining, manufacturing (also reported as a separate subgroup), construction, electricity, water, and gas. Source: World Bank WDI.
    \item Human capital index: It is a measure of the skills, education, capabilities, and attributes of the workforce that affect their productive capacity and potential earnings. Source: Penn World Tables.
    \item Renewable energy consumption (\% of total final energy consumption):	Renewable energy consumption is the share of renewable energy in total final energy consumption.	Source: World Bank WDI.
    \item Fossil fuel energy consumption (\% of total):	Fossil fuel comprises coal, oil, petroleum, and natural gas products. Source: World Bank WDI.
    \item Energy use (kg of oil equivalent per capita):	Energy use refers to use of primary energy before transformation to other end-use fuels, which is equal to indigenous production plus imports and stock changes, minus exports and fuels supplied to ships and aircraft engaged in international transport. Source: World Bank WDI.
    \item Alternative and nuclear energy (\% of total energy use): Clean energy is noncarbohydrate energy that does not produce carbon dioxide when generated. It includes hydropower and nuclear, geothermal, and solar power, among others. Source: World Bank WDI.
    \item Forest area (\% of land area):	Forest area is land under natural or planted stands of trees of at least 5 meters in situ, whether productive or not, and excludes tree stands in agricultural production systems (for example, in fruit plantations and agroforestry systems) and trees in urban parks and gardens. Source: World Bank WDI.
    \item Credit to the Private Sector (\% GDP):	As a proxy for the size of the domestic financial market. Source: World Bank.
    \item Chinn-Ito Index: a de jure measurement of the international financialization of a country. Source: \cite{chinnito2006}.
    \item KOF Financial Globalization Index:	KOF Globalization Index measures the economic, social and political dimensions of globalization. Source: \cite{dreher2006} and \cite{gygli2019}.
    \item KOF Trade Globalization Index:	KOF Globalization Index measures the economic, social and political dimensions of globalization. Source: \cite{dreher2006} and \cite{gygli2019}.
    \item International Interest Rate: Federal Funds Effective Rate. Source: Federal Reserve Economic Data.
    \item Net capital inflows (\% GDP):	Total financial liabilities minus total financial assets, excluding foreign exchange reserves, as \% of GDP. Source: IMF IFS.
    \item Real GDP Growth (annual \%): Annual percentage growth rate of GDP at market prices based on constant local currency. Source: World Bank WDI.
    \item Urban population (\% of total population):	Urban population refers to people living in urban areas as defined by national statistical offices. Source: World Bank WDI.
\end{itemize}    

\clearpage
\begin{table}[!htbp]
 \caption{Descriptive Statistics}\label{tab:stats}
   \justify
    \scalebox{0.6} {   
	  \setlength{\tabcolsep}{2pt} 
	  \estwide{stats.tex}{9}{c}
	  }
  \justify
 \scriptsize
 Notes: Own elaboration.
 \end{table}

\end{document}